\documentclass[11pt]{llncs}

\usepackage[T1]{fontenc}
\usepackage{tgbonum}

\usepackage{cite}

\usepackage{times}
\usepackage{epsf}
\usepackage{latexsym}
\usepackage{amsmath}
\usepackage{amsfonts}
\usepackage{euscript}
\usepackage{url}
\usepackage{graphicx}
\usepackage{cite}
\usepackage{subfigure}
\usepackage{enumitem}
\usepackage{authblk}
\usepackage{amssymb}
\usepackage{pifont}
\newcommand{\cmark}{\ding{51}}%
\newcommand{\xmark}{\ding{55}}%

\newcommand{\ignore}[1]{}

\begin{document}

\title{{UTXO} in Digital Currencies: Account-based or Token-based? Or Both?}
\author{Aldar C-F. Chan}

\institute{University of Hong Kong\\ aldar@graduate.hku.hk}
\maketitle

\begin{abstract}
There are different interpretations of the terms ``tokens'' and ``token-based systems'' in the literature around blockchain and digital currencies although the distinction between token-based and account-based systems is well entrenched in economics.  Despite the wide use of the terminologies of tokens and tokenisation in the cryptocurrency community, the underlying concept sometimes does not square well with the economic notions, or is even contrary to them.  The UTXO design of Bitcoin exhibits partially characteristics of a token-based system and partially characteristics of an account-based system.  A discussion on the difficulty to implement the economic notion of tokens in the digital domain, along with an exposition of the design of a UTXO, is given in order to discuss why UTXO-based systems should be viewed as account-based according to the classical economic notion.  Besides, a detailed comparison between UTXO-based systems and account-based systems is presented.  Using the data structure of the global/system state representation as the defining feature to distinguish digital token-based and account-based systems is therefore suggested.  This extended definition of token-based systems covers both physical and digital tokens while neatly distinguishing token-based and account-based systems.
\end{abstract}

\section{Introduction}
\label{sect::introduction}

The dichotomy of token-based and account-based models for payments systems is often discussed in the literature around blockchain and digital currencies \cite{AdrianM19, AuerB20, BIS18, BIS21, BordoL17, BrunnermeierJL21, Duffe19, GarrattLMM20, Group30, Kahn20, KahnRW20, Milne18, MPAAKPR18, LeeMW20}, including central bank digital currency (CBDC) and crypto-assets.  In the context of digital currencies, there are a range of different and contrasting interpretations of the terms ``tokens'' or ``token-based systems.''  In particular, the term ``tokens'' is often used to refer to designs or applications that are not necessarily directly linked to the concept of token-based systems.

Most drastically, the terminology regarding tokens in the cryptocurrency community, including the concept of tokenisation, usually refers to an implementation that is unambiguously an account-based system.
For example, the Ethereum community proposed a standard for fungible units of value termed ``tokens,'' which was introduced in the Ethereum Whitepaper \cite{Buterin13}.  The adopted standard, widely known by its proposal identifier ERC-20, is arguably the primary reference point for the concept of tokens on Ethereum and other public blockchain today.  Although called tokens, ERC-20 tokens are recorded in the form of an account balance (under an account address) in the smart contract copies hosted on replicated databases of the public blockchain, not stored as digital ``objects'' in the user's software wallet.  The wallet only store the private key used to sign instructions sent to the smart contract.  In essence, this is an account-based system.

For the case of Bitcoin \cite{Nakamoto08}, it exhibits the characteristics partially of token-based systems and partially of account-based systems.  Bitcoin handles record-keeping using a format known as ``unspent transaction outputs'' (more commonly referred to as UTXO), which are a data structure sharing many similarities with objects in a token-based system.  On the other hand, a Bitcoin address is in essence an account and the private key associated with the Bitcoin address is the proof of identity needed to transact from that account.  Funds are not stored as objects in user software wallets but recorded on a ledger (which is the public blockchain of Bitcoin), and the process of moving funds requires intermediation through the collective works of Bitcoin miners \cite{Milne18}.  Some researchers therefore hold the view that Bitcoin can be both token-based and account-based, and suggest that ``future classifications could modify the definitions of the terms account-based and token-based to more clearly distinguish them'' \cite{GarrattLMM20}.

This article discusses why UTXOs should be seen as an account-based arrangement according to the classical economic notions of tokens and accounts \cite{Green07, KahnR09, Kahn16}.  The distinction between token-based methods and account-based systems is pretty well entrenched in payment economics: For a transaction to be deemed satisfactory in an account-based system, the payer has to be identified as the holder of the account from which the payment will be made, whereas, in a token-based system, what needs to be identified is the genuineness of the object being transferred.  Through explaining the details of transaction processing in Bitcoin's UTXO, this article discusses why a UTXO does not fulfil this definition of token-based systems but is closer to the definition of account-based systems.  It also explains why achieving purely peer-to-peer, decentralised exchanges without any intermediation by third parties in the digital domain is difficult.  A comparison between UTXO-based systems and account-based systems is presented. Finally, a suggestion is made on the defining features of digital tokens (i.e. the data structure used to represent the system state) that may be used to create a new taxonomy suitable for distinguishing digital token-based systems from other arrangements.

The key contributions of this article is two-fold.  First, a detailed exposition of the design of a UTXO and a discussion on whether it should be classified as account-based systems are given with well-grounded justifications.  Second, an extension of the definition of token-based systems based on the global system state representation of the respective record system is proposed, which neatly distinguishes between token-based systems and account-based systems.  The resulting taxonomy could cover both physical and digital token-based systems.

This article is organised as follows.  Section~\ref{sect::economic_notion} discusses the classical distinction of tokens and accounts in payment economics.  Section~\ref{sect::difficulty_economic_tokens} why the economic notions of tokens is difficult to achieve in the digital domain.  Section~\ref{sect::bitcoin_utxo} presents the design of a UTXO and the transaction process of Bitcoin, highlighting the differences between UTXO-based and account-based implementations of blockchain.  A discussion on why UTXO should be viewed as account-based is given in Section~\ref{sect::taxonomy_problem}, with implications and suggestions for a new way to distinguish token-based systems and account-based systems given in Section~\ref{sect::taxonomy}.

\section{Economic Notions of Tokens and Accounts}
\label{sect::economic_notion}

The conceptual distinction between payments methods that are based on tokens (e.g. coins and banknotes) and payments systems that are based on accounts (e.g. bank deposits) is pretty well entrenched in discussions of payment economics \cite{Green07, KahnR09, Kahn16}.  As highlighted by \cite{Kahn20, BrunnermeierJL21}, the key difference between the two types of arrangements lies in the verification process for payments.  In an account-based system, the crucial question for the recipient is the payer's identity, whereas, in a token-based system, the payer's identity is irrelevant and, instead, the crucial question for the recipient is the authenticity of the received object  --- whether it is real or counterfeit.  The same dichotomy is commonly applied to the discussions of digital currencies and crypto-assets, including Bitcoin \cite{AdrianM19, AuerB20, BechG17, BordoL17, BrunnermeierJL21, Duffe19, GarrattLMM20, KahnRW20, LeeMW20, MPAAKPR18, Milne18}.

\subsection{Kocherlakota's Distinction between Money and Memory}
\label{ssect::Kocherlakota_distinction}
The discussion of the equivalence of money to a primitive form of memory in resource allocation \cite{Kocherlakota98} is sometimes seen as the origin of the distinction between accounts and tokens.  But \cite{Kocherlakota98} does not give any explicit definition for tokens.  To distinguish between money and memory, \cite{Kocherlakota98} defines money as ``an object that does not enter utility or production functions'' and calls it ``barren tokens'' allowing societies to achieve allocations that would otherwise not be achievable.  The characterization of tokens in \cite{Kocherlakota98} is limited.  In contrast, \cite{Kocherlakota98} has a much clearer exposition of account-based payment systems: ``An imaginary balance sheet is kept for each agent.  When an individual gives consumption to someone else, his balance rises, and his capacity for receiving future transfers goes up.  When he gets consumption from someone else, his balance falls, and his capacity for receiving future transfers declines.''

It should be noted that \cite{Kocherlakota98} actually considers a much broader concept than accounts, namely, ``memory.''  It defines memory as ``knowledge on the part of an agent of the full histories of all agents with whom he has had direct or indirect contact in the past.''  In fact, this definition precisely describes how the distributed ledger of Bitcoin works.  The ledger of Bitcoin, called blockchain, records all confirmed transactions that have ever occurred and are accessible to all participants.  If the memory of \cite{Kocherlakota98} is seen as an account-based system, so should Bitcoin \cite{Nakamoto08} be.  This is contrary to the common conception that Bitcoin and other crypto-assets are monetary tokens \cite{AdrianM19, KahnRW20, BordoL17, BrunnermeierJL21, Duffe19}.

\subsection{Kahn's Distinction between Tokens and Accounts}
\label{ssect::Kahn_distinction}
The most succinct distinction between accounts and tokens is presented by \cite{KahnR09}, which is widely adopted by discussions on digital currencies and crypto-assets \cite{AdrianM19, AuerB20, BechG17, BIS18, BordoL17, BrunnermeierJL21, Duffe19, GarrattLMM20, KahnRW20, LeeMW20, MPAAKPR18}.  It defines token-based systems as store-of-value systems (such as commodity money, fiat money, and stored value cards) that ``are founded on the transfer of some payments object (be it coins, notes, or electronic stored value) between payer and payee, and depend critically on a payee’s ability to verify the payment objects.''  As \cite{AdrianM19} highlights, ``a payment transaction is settled immediately as long as a payee deems the object to be valid, and no exchange of information is necessary.''
In contrast, \cite{KahnR09} describes account-based systems (such as charge accounts, checks, and credit cards) as record systems that ``require the keeping of accounts in the name of the payer and payee'' with their success hinging, most fundamentally on ``the ability of its participants to verify the identities of account holders, to ascertain the link between transactors and histories.''  These record systems record claims on values and, as \cite{AdrianM19} explains, ``a payment is made through a transfer of claim on value existing elsewhere'' and ``a payment therefore requires giving instruction to transfer ownership of a claim on the value from one person to another.''
In other words, in order to sufficiently verify the validity of a payment transaction, a token-based system requires verifying the validity of the object used to pay, whereas, an account-based system requires verifying the identity of the payer.\footnote{Of course, for an account-based system, there are also other checks (such as whether the payer's account have a sufficient balance to settle the transaction) in order to successfully process a transaction.}

\subsection{Canonical Examples: Cash versus Bank Deposits}
As summarized in Table~\ref{table:token_versus_account}, the differences between tokens and accounts could be better illustrated through the canonical examples of cash (for token-based methods) and bank deposits (for account-based systems).
Both token-based and account-based systems are record-keeping arrangements to record the ownerships of an asset \cite{KahnRW20}, which is cash or a bank deposit in our discussion.  {\em The difference between them lies in how the records are kept, and who keeps and updates the records.}  In short, in a token-based system, the record is kept and updated by the owners of tokens, whereas, in an account-based system, the record is usually kept and updated by an operator of a system infrastructure playing the role of a trusted intermediary.

\begin{table}[hbpt]
    \center{
    \begin{tabular}{|l|c|c|}
      \hline
       & Bank deposits as an embodiment  & Cash as an embodiment \\
       & of account-based systems        & of token-based systems\\
       \hline
      Organisation of system records & account-oriented & object-oriented \\
                        & (recorded as a list of  & (recorded as a list of  \\
                        & accounts and balances) & assets and ownership) \\
      \hline
      Intermediary required to record ownerships & \cmark & \xmark \\
      \hline
      Peer-to-peer exchange of asset& \xmark & \cmark \\
      \hline
      Payer's identity required to initiate payments & \cmark & \xmark \\
      \hline
      Knowledge of payee's identity required for & \cmark & \xmark \\
      payment transactions                       &     &    \\
      \hline
      Fungible units of value & \cmark & \xmark \\
      \hline
      Value aggregation & \cmark & \xmark \\
      \hline
    \end{tabular}
    }
    \vskip 0.5cm

    \caption{Differences between token-based and account-based systems as illustrated by cash and bank deposits.}
    \label{table:token_versus_account}
\end{table}

In a token-based system, the payments object (i.e. cash) is the record summarizing past production, trade and consumption decisions \cite{KahnRW20}.  The possession of the object is generally viewed as a proof of ownership of cash since each object has a unique physical existence (i.e. the same object can only be owned by one entity at any particular instant of time).  Once the payer hands it over to the payee, he no longer owns it and, instead, the payee owns it.  The genuineness of the object being transferred is all that a payee needs to verify in order to confirm the validity of the transfer of ownership.  No trusted party is required to attest this transfer of ownership.  The payee assumes all the liability if the cash is counterfeit.  Nor is an intermediary  required to keep the records of ownerships.  As cash changes hands, the change in possession of the object amounts to updating the records in the system.  The exchange is decentralised or peer-to-peer (i.e. without relying on intermediaries for updating ownership records).  Besides, the payer is not required to know the identity of the payee as the change in possession of the object occurs.  Nor is the identity of the payer needed to corroborate the ownership of the object.  Hence, the payment is {\em anonymous}.  This record system is decentralized because there is no single source of the records and no single party responsible for updating the records.  Most of the crypto-assets such as Bitcoin also have their records of ownerships distributed in a network of nodes and updated in a decentralised fashion \cite{Nakamoto08}.  The difference from cash lies in whether third parties are relied on to update the records.  Transfers of ownership are also anonymous on these networks, and the payee of a transaction assumes all responsibilities if fraud occurs.  As a result, many tend to see Bitcoin and other crypto-assets as tokens.  However, intermediaries are needed in Bitcoin or Ethereum to record ownerships and update records to process changes in ownership, which is different from standard token-based systems.

In contrast, the records of ownerships in an account-based system like bank deposits are usually centralised, entrusted to and updated by a single party or intermediary, that is, the operator of the system infrastructure.  This intermediary is usually the party on which the claims of bank deposits are.  The intermediary is required to keep all records of ownerships and update them as payments are made to effect transfers of ownerships.  Usually, these records are maintained in the form of account balances under the names of the owners, and a transfer of ownership therefore comprises of a debit of the account balance of the payer and a credit of the account balance of the payee.  As a result, the intermediary would need to verify the identity of the payer of a transaction in order to ensure that the payment is a properly authorised instruction from the account holder.  If it is later discovered that the intermediary has incorrectly identified the payer, it assumes liability and would need to refund the account holder.  Besides, the payee's identity is also needed in a payment transaction in order to credit the payment to the right account.  It should be noted that it is usually the account numbers, rather than real personally identifiable information of account owners, that are used in processing payment transactions.  Whether payments can be anonymous depends on the requirement of the intermediary recording account entries and processing payments.

Another subtle difference between token-based and account-based systems is the record or data structure used for accounting.  In computer science terms, the global or system state representations of the two systems are different.  A token-based system like cash records the state of the system as a list of individual assets or objects (i.e. tokens) each of which has a corresponding owner (whose real identity may or may not be known\footnote{Note that the record system does not exist physically for a token-based system and is merely imaginary.  Taking cash as an example, the record system is instantiated through each owner physically possessing the respective payment object.  Besides, the real identity of the owner is not publicly known.}) who can spend the asset.  Each of these tokens has a specific value, which does not change.  When two tokens, such as banknotes, of the same face values are received from different payers, their values would not be aggregated and the tokens remain distinguishable from each other, namely, that the payee can still tell apart the tokens received from different payers (say, through the serial numbers of the banknotes).  That is, tokens are non-fungible or uniquely distinguishable from one another.
By contrast, an account-based system records the state of the system as a list of accounts, each of which is under an account holder and has a corresponding balance.  When funds are transferred, the system record is updated by increasing and decreasing the balances in the relevant account.  When two payments are received by an account, their transaction values would be added up with the existing balance to update the system record with a new balance.  The values of the two payments are subsumed by the new account balance, and nobody can tell from the account balance whether it is a result wholly from one single payment or from multiple payments previously received.  For Bitcoin, no aggregation occurs, whereas, for crypto-assets on Ethereum, aggregation occurs for each defined asset.

\subsection{Formal Definition of Record System}
\label{ssect:formal_definition}

A formal definition of the data structure of the record system could help understand the key distinction between token-based systems and account-based systems.  This also helps formulating a framework for digital tokens in Section~\ref{sect::taxonomy_problem}.  Let ${\cal U}$ denote the set of all users of a system, which can be either token-based or account-based, ${\cal O}$ denote the set of all payment objects issued in a token-based system, and $\mathbb{R}$ the set of real numbers.

The global state ${\cal S}_{t}^{account}$  of the record system of an account-based system at any time instant $t$ is then given by:
\[
{\cal S}_{t}^{account} = \{(u_i, b_i): u_i \in {\cal U}, b_i \in \mathbb{R}\}
\]
where $b_i$ is the account balance of user $u_i$.

A transaction is a 3-tuple $T = (u_i, u_j, x)$, where a user $u_i$ is the payer and pays another user $u_j$, the payee, with $x$ units of value at time $t$, to process the transaction $T(u_i, u_i, x)$, the global state is updated as follows:
\[
{\cal S}_{t+1}^{account} = \{{\cal S}_{t}^{account} \backslash\ \{(u_i, b_i), (u_j, b_j)\}\} \bigcup \{(u_i, b_i-x), (u_j, b_j+x)\}
\]

Note that the account balance of $u_i$ is debited and that of $u_j$is credited.

By contrast, the global state ${\cal S}_t^{token}$ of the record system of a token-based system at any time instant $t$ is given by:
\[
{\cal S}_t^{token} =\{(o_k, u_i): o_k \in {\cal O}, u_i \in {\cal U}\}
\]

A transaction for a token-based system is also a 3-tuple $T=(u_i, u_j, o_k)$.  To process a transaction $T(u_i, u_j, o_k)$ which user $u_i$ pays another user $u_j$ a payment object (say, a banknote) $o_k$, the global state is updated as follows:
\[
{\cal S}_{t+1}^{token} = \{{\cal S}_{t}^{token} \backslash \{(o_k, u_i)\}\} \bigcup \{(o_k, u_j)\}
\]
Note that the value of the payment $o_k$ stays constant with its ownership transferred from $u_i$ to $u_j$.  For the case of cash, the value of a banknote stays the same but the ownership changes as exchange occurs.

The record system of a token-based system is an imaginary system which does not exist physically, whereas, the record system of an account-based system has physical existence.  The storage or memory size of an account-based system is $|\{u_i\}|$ which is proportional to the total number of accounts recorded in the system, whereas, the size of records for a token-based system (which is imaginary) is $|\{o_k\}|$, proportional to the total number of payment objects ever issued.

\section{Difficulty to Implement Digital Tokens in the Economic Sense}
\label{sect::difficulty_economic_tokens}

A salient characteristic of the economic notion of a token is the sufficiency of verifying the genuineness of an object (the token) to confirm the validity of a payment or transfer of ownership of the token, which stands in sharp contrast to an account-based system requiring an intermediary to record and update account balances.  This sufficiency condition enables decentralised, peer-to-peer settlement of payments without relying on any intermediary in a pure token-based system.  This capability is attributed to the fact that a token has a unique physical existence.  Implementing this capability in a purely digital domain is challenging for a number of reasons, including the ease of creating an exact replica of any piece of digital data, lack of recipient particularity in the digital domain and possibility of double spending by payers.

The transfer of ownership of a physical token can be easily and promptly confirmed because of its unique physical existence.  When a payer transfers a physical token to a payee, he no longer possesses it and there is no ambiguity of the ownership of the token.  However, a digital token is merely a piece of digital data.  Like any digital data, it can be easily copied and replicated, virtually at no cost, implying that a ``unique'' virtual existence of a digital token does not exist.  On one hand, after passing a digital token to a payee, the payer would still have a copy of it.  The ownership of the token cannot be simply proven through the possession of a copy of the token; the transfer of ownership of a digital token is not settled as neatly as that of a physical token.  Besides, a plain transfer of a digital token lacks recipient particularity.\footnote{Unless a token includes explicit information of the recipient and is signed by the payer, its ownership cannot be easily verified.}  Without a physical dimension or contextual factor, it is impossible to prevent others from receiving the same digital object and thereby claiming its ownership \cite{ChanZ14}.  Anyone possessing a copy of a digital token can actually claim its ownership.  When presented with multiple, exact replicas of a digital token, it is non-trivial for anyone to determine which of these replicas or copies is actually original and carries or inherits the underlying (monetary) value of the asset represented by the token.

\subsection{The Bitcoin Approach}
\label{ssect::approach_of_bitcoin}

To solve the ownership ambiguity issue of digital tokens, a Bitcoin transaction usually includes the payee's identity to specify the designated recipient of a payment.  A Pay-to-Public-Key-Hash (P2PKH) transaction --- the most common form of Bitcoin transactions --- includes (the hash of) the public key of payee and the amount to be paid to this payee in a transaction format called UTXO.  \cite{NarayananBFMG16} rightly points out that the public key (or its hash) is a form of identity for the payee.  In this way, the value transfer is explicitly directed to the desired payee unambiguously, and the rightful ownership of a UTXO can be correctly corroborated by looking at the specified public key hash.

However, explicitly specifying the payee's identity in a payment transaction does not suffice to prevent all kinds of fraud.  It cannot prevent an owner of bitcoins from creating multiple distinct transactions to pay different payees with the same UTXO or bitcoins.  This is the ``double spending'' problem \cite{Nakamoto08}.  After a payer spends a given UTXO or bitcoin, subsequent transactions spending the same UTXO should not be accepted since the ownership of the value underlining the UTXO has been transferred to the payee and the payer is no longer the rightful owner.  However, without a trusted intermediary to keep track of the temporal order of transactions, it is difficult for anyone (except the first payee who can see the same UTXO spent twice if presented with the new transaction) to tell whether the payer still owns the UTXO or bitcoin.  Intermediation by a trusted party is necessary to ensure the uniqueness of current ownerships of bitcoins.  The job of the intermediary is to check whether the UTXO has been spent previously.

In Bitcoin, a system-wide intermediary or timestamping service is implemented over a peer-to-peer network of nodes in a distributed fashion to record the current, rightful ownerships of all UTXOs or bitcoins in a ledger which is collectively updated by the participating nodes through the PoW (Proof of Work) process.  As \cite{Milne18} argues, Bitcoin would be better labelled as account-based since it records ownerships of coins (more precisely, UTXOs) on a ledger and any transfer of ownership requires a signed transaction from the owner in order for the blockchain to update the record.  The purpose of this signed transaction is similar to that of a signed instruction from the owner for the case of bank deposits.

\begin{figure}[htbp]
    \begin{center}
    \includegraphics[width=18cm]{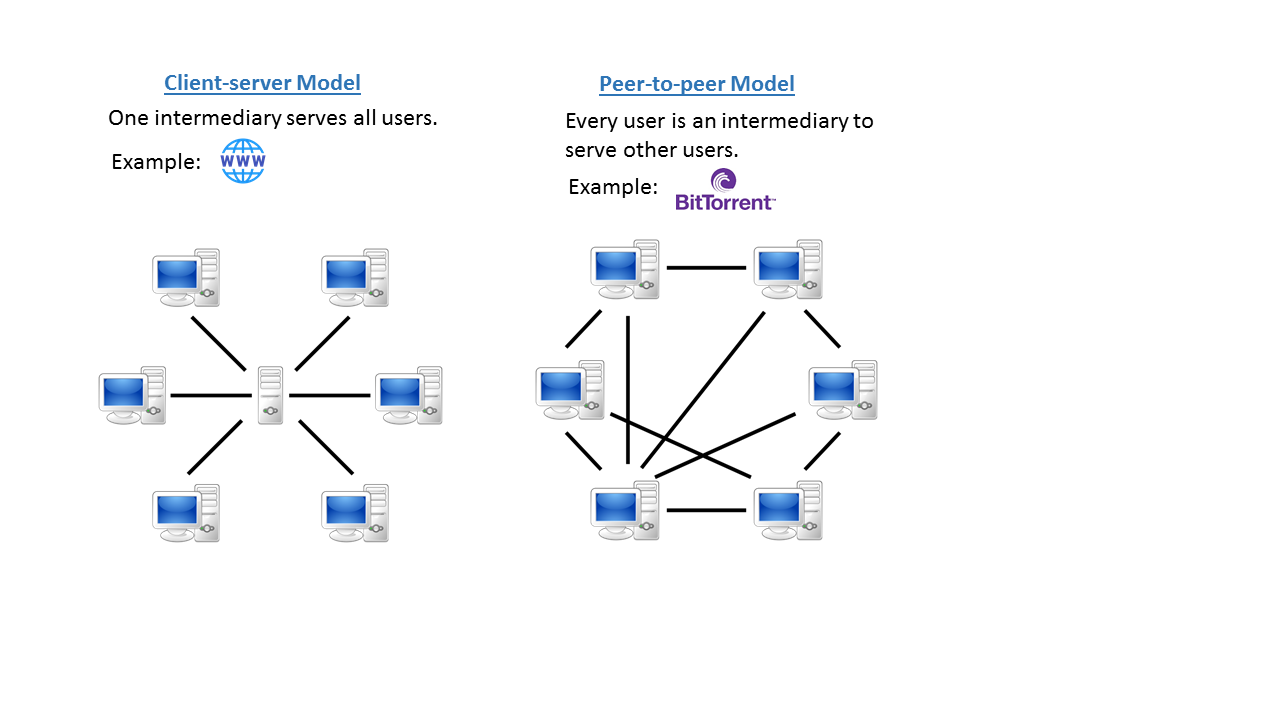}
    \end{center}
    \vskip -2.5cm
    \caption{Differences between Client-Server Model and Peer-to-peer Model.}
    \label{figure:client-server-p2p}
\end{figure}

A common confusion is that Bitcoin, which claims to be a peer-to-peer payment system, is seen as supporting peer-to-peer exchange of value without the need of any intermediation.
The peer-to-peer setting in Bitcoin refers to a model in computer science which is in contrast to the typical client-server model (Figure~\ref{figure:client-server-p2p}), as the Bitcoin whitepaper \cite{Nakamoto08} puts it: ``... we propose a solution to the double-spending problem using a {\em peer-to-peer distributed timestamp server} to generate computational proof of the chronological order of transactions.''  That is, the peer-to-peer nature clearly refers to the implementation of the timestamping service, rather than the way of transfer of ownerships or an asset or payments.  This is distinctly different from the economic notion of peer-to-peer payments in a token-based system, which do not require any intermediation.  Yet, it is still reasonable to say that the transfer of value in Bitcoin is peer-to-peer because, like typical token-based systems such as cash, the transacting parties do not need to have established trust on the intermediaries (or miners) which updates the ownership records \cite{BechG17}.\footnote{Although the transacting parties does not have to trust a single miner in the Bitcoin network, they have to trust that the majority of miners in the network follow the protocol as implemented in the code they are running.}
In the light of the payment process and record booking of Bitcoin, {\em it is more appropriate to see Bitcoin as account-based, rather than token-based in the economic sense as payments involve verifying and updating accounting entries (in the form of UTXO representations) on a blockchain or distributed ledger.}

\subsection{The Approach of Chaum's e-cash}
\label{ssect::chaum_e_cash}
Chaum's e-cash \cite{Chaum83} is closer to a token-based system.  The proof of ownership of a digital token is truly based on the possession of a signed message of token information.  The tradeoff is that each coin can only used once after issuance and has to be redeemed at the issuer immediately upon receipt.  In the scheme, a unique public-private key pair is chosen for each denomination of coins and made known publicly.  Using a blind signature scheme, a payer can request the coin issuer to sign on a coin while staying anonymously.  The message to be signed for a coin could simply be its serial number and the nominal value of the coin is matched with the signing key of the issuer.  The payer can then strip off the blinding factor to obtain a coin which, in essence, is a signed message containing a serial number signed by a public key of the issuer, more precisely, the public key corresponding to the denomination of the coin.  Anyone can verify the authenticity of the coin by checking the issuer's signature and its value can be attested the signature can be verified correctly with the corresponding public key.  The payer could pay the coin to a payee by simply sending the message signed by the issuer.  To confirm the validity of the payment, the payee has to deposit the coin immediately back to the issuer, which keeps a list of used coins.  The payment is valid only if the issuer confirms that the coin has not been used.  In other words, a ledger is kept by the issuer and required to keep track of the consumption status of all coins.  This can possibly be seen as closer to a token-based system since entries in the ledger are coin serial numbers only and no identity information of users is kept by the issuer's ledger.

\bigskip
As can be seen in the examples of Bitcoin and Chaum's e-cash, a record system is inevitable for the implementation of digital tokens.  To achieve the decentralised, peer-to-peer exchange of a digital token, a physical factor needs to be included.  The use of trusted computing, such as the secure element of a smart card, ARM's Trustzone and Intel's SGX, may help reduce the reliance on an intermediary for keeping and updating ownership records.  However, a trusted party is still needed for bootstrapping such systems.

\section{The UTXO Implementation of Bitcoin}
\label{sect::bitcoin_utxo}
A Bitcoin user software wallet (i.e. the mobile app) stores a collection of public-private key pairs of a user.  It does not store bitcoins directly, which is in sharp contrast to typical token-based systems, which store payment objects in user wallets.  Instead, the bitcoin holding of the user is recorded on the Bitcoin blockchain (which can be seen as a ledger), with multiple copies of this ledger kept and updated by different nodes in the Bitcoin network.  A Bitcoin address is an identifier of 26–35 alphanumeric characters, which is strictly derived from the hash of a public key of the user.  The Bitcoin address has a similar function to account numbers in a typical account-based system, and is used to hold bitcoins.  That is, the information of bitcoin holding of a Bitcoin address is recorded on the blockchain, but with a different data structure when compared with a typical account-based system.

In contrast to a typical account-based system (which records the system state as a list of accounts, each with a corresponding balance), the Bitcoin blockchain records the system state as a list of individual bitcoins (more precisely, UTXOs), each of which has a corresponding owner who can spend the bitcoins.  Each of the UTXO has a specific value.  In order to initiate a transfer, the holder of a UTXO is required to prove that they own it by signing a payment transaction (which in essence is an instruction) with the private key associated with the UTXO record on the blockchain.  An individual UTXO cannot be partially spent; instead, the UTXO being transferred is generally destroyed and replaced with two newly created smaller UTXOs (which sum up to the same total value), with one going to the payee and the other being returned to the payer as change.  That is, each UTXO currently active on the blockchain inherits its value from one or more UTXOs which it consumes.  The only exception is the UTXO of a coinbase transaction which records newly minted bitcoins.  A coinbase UTXO is valid if it is included in one of the blocks that have been admitted into the Bitcoin.

A Bitcoin transaction is composed of UTXOs, which represent bitcoins received by a user in a previous transaction.  Except coinbase transactions (with newly minted bitcoins which are created in each block by mining), each transaction has inputs and outputs.  A transaction input includes a reference to bitcoins received in a previous transaction (i.e. represented by a UTXO) and a proof that these bitcoins or UTXOs belongs to the payer who wants to spend them.  For example, if Alice wants to send 5 bitcoins (BTCs) to Bob, she needs to refer to other transactions which she has previously received and whose amount is at least 5 bitcoins.  A transaction output describes the destination of bitcoins by providing a locking script.  The locking script is a challenge which the payee needs to solve in order to spend the bitcoins in a subsequent transaction, and can be seen as the conditions for spending the concerned bitcoins.  To prove the ownership of the bitcoins in order to spend them in a new transaction, the payee needs to present in a transaction input an unlocking script which matches the locking script of the output of a previous transaction containing the bitcoins.  A fairy wide range of conditions can be specified in the locking/unlocking scripts of Bitcoin.  The unlocking script, augmented with the locking script, will be executed by nodes or miners of the Bitcoin network to check the validity of the transaction.  If the execution of the script produces a result of TRUE, the transaction is deemed as valid and included in the blockchain after the consensus among the miners completes.

Since each transaction has a reference to previous transactions in order to inherit their value, transactions can be linked up together to form an inheritance chain from any currently active UTXO to the coinbase UTXO.  The sequential verification of the digital signature of transactions in this chain, along with the fact that the concerned coinbase transaction can be found in one of the admitted blocks of the blockchain, can attest the authenticity of the current UTXO.  That is why some researchers see this verification of an inheritance chain as the counterpart of the deed of verifying the genuineness of a payment object in typical token-based systems \cite{GarrattLMM20}.  However, it should be noted that, in the real implementation of Bitcoin, this verification process is not necessary.  Instead, a database of active UTXOs are maintained by each miner of Bitcoin.  When a UTXO is consumed by a transaction, it will be removed from the database and the UTXOs of the transaction will be added to replace it.  This implementation actually renders Bitcoin resemble an account-based system more closely.

\subsection{Pay-to-Public-Key-Hash (P2PKH) Transactions}
\label{ssect::p2pkh}

Pay-to-Public-Key-Hash (P2PKH) transactions are the most common transactions in Bitcoin.  These transactions contain a locking script that encumbers the output with a public key hash.  They requires the proof of ownership of the public key in order to spend the bitcoins.  A P2PKH output can be unlocked (spent) by a public key and a digital signature created with the corresponding private key.  Blockchain nodes would verify the correctness of this signature against the public key in the locking script.  In the light of public keys as identities \cite{NarayananBFMG16}, creating a digital signature in the unlocking script to spend the bitcoins would be equivalent to proving the identity to blockchain nodes in order to request them to update ownership records as the payment is processed.  In other words, the operation of P2PKH transactions is like a typical account-based system.

A P2PKH transaction fulfills the account-based system definition of \cite{AuerB20, BIS21} that an account-based system ``ties ownership to an identity'' in such a way that ``claims are represented in a database that records the value along with a reference to the identity'' and ''transactions are authorised via identification.''  The Bitcoin blockchain can be seen as a distributed version of the referred database to record bitcoins (value) and their ownerships as referred by the public keys (identities) of the owners.  \cite{AuerB20} considers signing a message with a private key for verification by others using the corresponding public key as a proof of knowledge of a secret (and therefore classifies the method as token-based).  However, creating a digital signature with the respective private keys to initiate a payment in Bitcoin can be better viewed as an identification process instead because this is a common embodiment of the well-known authentication factor ``something you know'' that has been widely used for entity authentication, along with ``something you have'' and ``something you are'' \cite{NIST20}, and public keys are generally viewed as identities \cite{NarayananBFMG16} in cryptography.

\subsection{Pay-to-Hash (P2H) Transactions}
\label{ssect:p2h}

Pay-to-Hash (P2H)  transactions (or the hash-lock puzzles) of Bitcoin are non-standard Bitcoin transactions.  These transactions contain a locking script which is a hash of an unknown hexadecimal string.  Anyone who can present a string which produces a correct hash value (i.e. the same as the locking script) would be able to spend the bitcoins.  A P2H transaction cannot be considered as a Hash Time-locked Contract (HTLC) since a P2H transaction has no expiration time while a HTLC has a time limit for a payee to accept the payment.  In other words, for P2H transactions, no identity is attached to a UTXO on the ledger record of the blockchain.  Instead of proving the identity by creating a respective digital signature, the owner of the UTXO needs to demonstrate knowledge of a secret (i.e. the string which produces a correct hash value matching the locking script).

P2H transactions fulfill the token definition of \cite{AuerB20, BIS21} and are in sharp contrast to typical account-based systems.  According to \cite{AuerB20, BIS21}, in a token-based system, ``claims are honoured based solely on demonstrated knowledge of an encrypted value."  For the case of P2H transactions, the hash value specified in a UTXO could be seen as an encrypted value and the owner needs to demonstrate knowledge of the preimage corresponding to this hash value in order to spend the UTXO.  However, it is also appropriate to see this UTXO record as a one-time account to receive bitcoins and then spend them, and therefore classify the design of P2H transactions as an account-based system.  In particular, P2H transactions are non-standard and only make up a very small percentage of Bitcoin transactions.

\section{Is Bitcoin Token-based?}
\label{sect::taxonomy_problem}

There is a stratified view on whether Bitcoin is token-based or account-based.  While Bitcoin is widely seen as a token-based system because of the perception that it supports decentralised, peer-to-peer exchange via blockchain, \cite{Milne18} rightly argues that Bitcoin should be classified as an account-based system for two reasons.  First, Bitcoins are not objects, which is a critical element of the economic notion of tokens \cite{KahnR09}. The account number in Bitcoin is the public key under which bitcoins (more, precisely, of P2PKH transactions) are held.  Second, transfer of bitcoins depends, just as with any other electronic transfer of an account-based money, on compliance with the transfer protocols of a payment scheme, which requires such an infrastructure of hardware, software, standards and supporting financial arrangements.

Bitcoin indeed fits the definition of an account-based system \cite{KahnR09, Kahn16}.  First, Bitcoin is truly a record system keeping accounts in the name of payers and payees as represented by their public keys.  As argued by \cite{GarrattLMM20, Milne18}, these public keys are account numbers in Bitcoin.  The difference is that, instead of recording a single balance under an account number, Bitcoin keeps multiple separate UTXOs under an account number.  Second, when someone wants to spend a bitcoin or UTXO, the protocol verifies a signed transaction (in essence, an instruction) from him with his public key.  In the light of public keys as identities \cite{NarayananBFMG16}, this signature generation could be seen as his proof of identity to the blockchain miners which can then ascertain the link between the transactor and histories (i.e. the ownership records on blockchain linked to his public key) before processing his transaction.

A possible argument against this view is that the public keys are not linked to real identities and transfer of bitcoin ownership is carried out anonymously.  In such an argument, signing a transaction is merely seen as a proof of knowledge, rather than a proof of identity.
For example, \cite{BrunnermeierJL21} states that ``to transact cryptocurrency, the payer must sign transactions with a “private key” linked to a particular set of coins, but the transaction is valid regardless of who presents that key,'' and no one is required to verify that the person who signed the transaction.  Along the same line of thought, the definition of a token-based system in \cite{AuerB20, BIS21} requires a payer to prove knowledge of a secret (e.g. through digital signature generation) in order to make payments.  The problem with this definition is that it does not create mutually exclusive categories in the taxonomy as typical account-based systems, such as bank accounts and RTGS (Real Time Gross Settlement) systems, may also fit this definition.  In an RTGS system, a bank initiates a payment through sending a signed instruction message such as MT202 messages.  A user presents his password to access his online bank account; this is a proof of knowledge of the password.  In other words, RTGS system and online bank accounts, which are widely accepted as examples of account-based systems and fit the definition of an account-based system, are also classified as token-based systems in accordance with the definition of \cite{AuerB20, BIS21, BrunnermeierJL21}.  Besides, remote entity authentication (or simply identification) is commonly based on one or more of the three authentication factors, namely, ``something you know'', ``something you have'' and ``something you are'' \cite{NIST20}.  Generating digital signatures is a common embodiment of the ``something you know'' factor for identification, a way for generating a proof of identity online.  That is, presenting a signature or a password through a challenge-response protocol is widely considered as a proof of online identities.

\begin{figure}
  \centering
  \includegraphics[width=18cm]{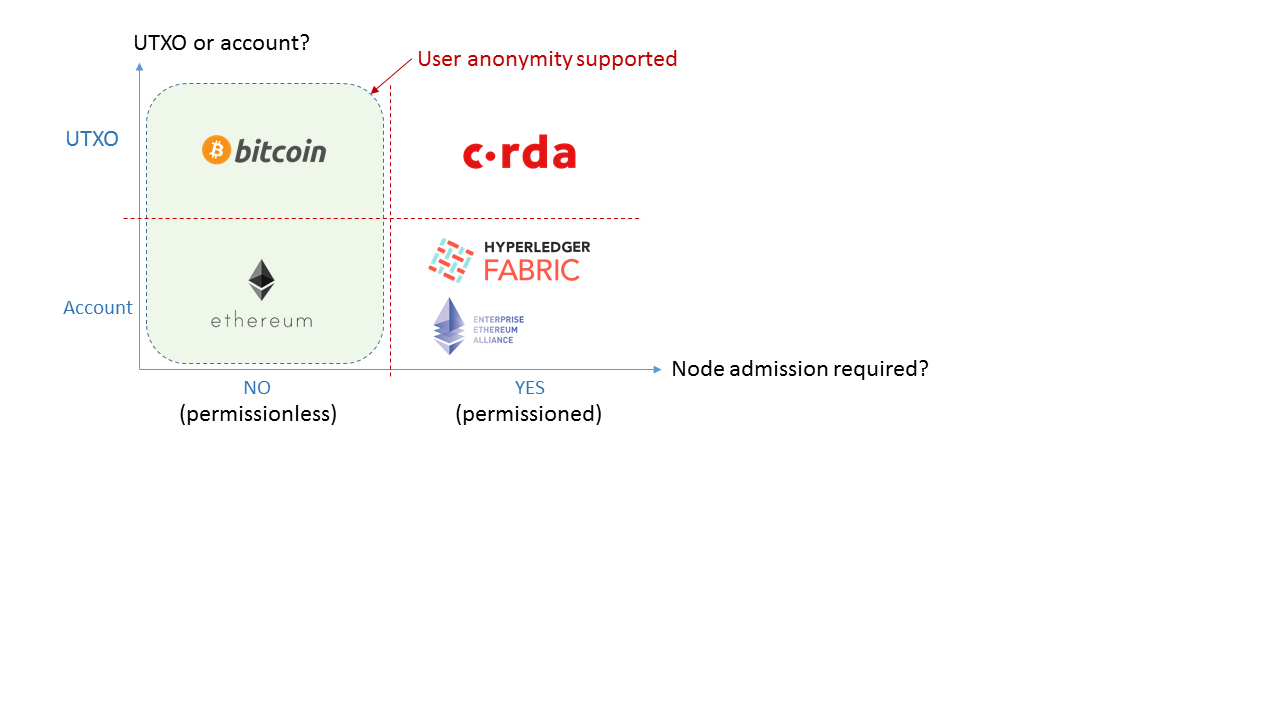}
  \vskip -4cm
  \caption{A comparison of anonymity property of account-based and UTXO-based systems.}\label{figure:anonymity}
\end{figure}

In addition, whether the real identity of the owner of a public key should be verified is a system design choice, rather than an inherent property of the UTXO design.  An account-based system like Ethereum also supports anonymous transactions.  This is the difference between the physical world and the cyber domain enable by cryptography.  As shown in Figure~\ref{figure:anonymity}, anonymity can be achieved by either UTXO-based or account-based systems, and UTXO-based systems may be designed to preclude anonymity in the system.  That is, supporting anonymity and UTXO designs are two independent design questions.  Anonymity should not be a defining feature for defining token-based systems in the digital domain.  As \cite{GarrattLMM20} argues, ``it is not relevant whether the system requires users to reveal their true identity," and ``rather, what matters is whether a user must follow a process the system has developed for verifying the identity that they established within the system, whatever that may be.''  The P2H transaction could possibly been seen as tokens according to the definition of \cite{AuerB20} as the owner of a bitcoin proves his ownership through proving knowledge of the preimage of a given hash value.  However, it is also possible to view this arrangement as a one-time account and the function of the preimage is similar to that of a password to access the account.

Besides seeing Bitcoin fit the definition of an account-based system, \cite{GarrattLMM20} also argues that Bitcoin fits the definition of a token-based system in the sense of the classical definition \cite{KahnR09}.  Its argument is that ``when someone wants to spend a bitcoin, the protocol verifies its validity by tracing its history,'' and ``the current transaction history is used to verify the validity of the “object” being transferred, as other token-based systems also do.''  It sees that, ``with Bitcoin, the object is a UTXO, which is only valid if it has not already been spent.''  However, this argument has a number of problems.  First, tracing the history of a bitcoin is to verify who is its rightful owner then, rather than whether it is genuine.  Second, a correct inheritance chain of the UTXO is not a {\em sufficient} condition for ascertaining its rightful owner due to the possibility of double spending.  A payee has to query the blockchain to check if its transaction has been included in one of the blocks; in fact, he has to wait for at least another six blocks after its transaction is included before being sure of the validity of the transfer.  This does not fit the definition of \cite{KahnR09} which implies that checking the genuineness of the payment object suffices as it needs to verify something more and rely on third-party intermediation.  Finally, in the real implementation of Bitcoin, the tracing of the history of a UTXO is not necessary either.  In practice, each full node maintains a UTXO database called the chainstate (introduced in 2012) to keep track of unspent UTXOs \cite{Bitcoin12}.  If the UTXO to be consumed is found in the database, it is seen as unspent and the owner’s signature on the transaction suffices to effect the payment.  This is analogous to checking whether the payer’s account balance is enough to make the payment in an account-based system.  A possible counter-argument is that the checking of the UTXO datatbase can be seen as verifying the genuineness of a UTXO.  However, the key question is what role a payee plays in checking the UTXO database.  Does he do it as the payee or as one of the operators of the blockchain?  Not all users install a full Bitcoin node in practice.

In summary, in the light of the need of a record system and intermediation for exchange and that public keys can be viewed as identities, a UTXO-based system should be viewed as an account-based system according to the classical economic definition which distinguish between token-based systems and account-based systems \cite{KahnR09, Kahn16}.

\subsection{Implications}
\label{ssect::implications}
That UTXO-based systems should be viewed as account-based have a number of implications.  First, neither an account-based system not a UTXO-based system would enable cash-like, peer-to-peer transfers, where a payment can be made without reference to any third party intermediary.  In account-based system, the accounts of the payer and payee need to be debited and credited by the operator of the ledger.  Whereas, in a UTXO-based system, to prevent double-spending, ownership of tokens needs to be recorded in a ledger, which will need to be updated to reflect any change in ownership.  Consequently, cross-network interoperability of UTXO-based systems are not as simple cash.  For account-based systems, a common balance sheet is needed for transfer between two separate systems, which can be provided by a common participant on both systems or by a third system that the two systems are participants of.  For the case of UTXO-based systems, UTXOs created in one system are not immediately portable to another system without setup.  Synchronisation between the record systems is inevitable.

Second, certain concerns with a digital token may not be material.  UTXO-based systems would not automatically provide anonymity.  In fact, both account-based systems and UTXO-based systems can be configured with various identity solutions, ranging from fully anonymous to pseudonymous and to a fully transparent, identifiable solution despite the complexity to implement the solution.  In general, implementing anonymity protection solutions on a UTXO-based system (or digital token systems such Chaum's e-cash \cite{Chaum83}) is relatively simpler, when compared with an account-based system.  For example, to implement a pseudonym system on an account-based system, the total number of active accounts, and hence the size of the global state of the system, could grow considerably.

Table~\ref{table:token_account_utxo} summarize the main differences between token-based, account-based and UTXO-based systems on different design issues, including cross-system interoperability, transaction traceability and complexity to implement transaction anonymity.  A detailed discussion is given in Section~\ref{sect::taxonomy} to explain why UTXO-based systems support transaction traceability while pure token-based systems do not.

\begin{table}
  \centering
  \begin{tabular}{|l|c|c|c|}
    \hline
     & Account-based & UTXO-based & Token-based \\
     & (e.g. Ethereum) & (e..g. Bitcoin) & (e.g. Chaum's e-cash)\\
     \hline
    Cross-system portability and synchronisation & difficult & easy & easy \\
    \hline
    Transaction traceability & no & yes & no \\
    \hline
    Complexity to implement transaction anonymity & high & low & low \\
    \hline
  \end{tabular}
  \vskip 0.5cm
  \caption{A simple comparison on design issues between token-based, account-based and UTXO-based systems}\label{table:token_account_utxo}
\end{table}

\section{Is a New Taxonomy Needed?}
\label{sect::taxonomy}

While UTXO-based systems are account-based systems, there are contrasting difference between the two arrangements attributed to the different ways of representations of the global or system state.  Summarized in Table~\ref{table:utxo_versus_account} is a comparison between UTXO-based systems and account-based systems.  Most of these differences result from the difference in representations of the state of the system.

\subsection{Differences between UTXO-based and account-based systems.}

In a UTXO-based implementation, the system state is represented by a list of ownerships of all units of value (or assets), and a transaction explicitly specifies the resulting state (i.e. ownerships) of the concerned units of value.  In this way, the right to issue a new unit of value is a prerogative of the respective asset issuer (as assured by public key cryptography), and the intermediary can only update the ownership of an existing, issued unit of value, but has no capability to issue new units.

In an account-based implementation, the system state is represented by a list of accounts and their balances, and a transaction does not carry full information that can allow the determination of the end state or resulting account balances of the involved holders.  A transaction has to be combined with the current state or concerned account balances in order to determine the resulting account balances.  In other words, repeatedly replaying a transaction could possibly lead to wrong account balances.

\begin{table}[hbpt]
    \center{
    \begin{tabular}{|l|l|l|}
      \hline
       & UTXO-based implementations & Account-based implementations\\
       & (e.g. Bitcoin, Corda) & (e.g. Ethereum, Hyperledger Fabric)\\
       \hline
      Global/system state & a list of assets or objects (UTXOs) & a list of accounts and their balances \\
      representation & and their ownership identifiers  & \\
      \hline
      Size of database & proportional to the number of & proportional to the number of \\
      & units of asset  & accounts \\
      \hline
      Information in a transaction & a transaction explicitly specifies  & a transaction needs to be combined  \\
      & the end state to be arrived & with the current system state in order \\
      & & to determine the end state to be \\
      & & arrived\\
      \hline
      How a payment transaction  & updating the ownership of an asset  & debiting the payer's account balance  \\
      is processed & from the payer's identifier to the   & and crediting the payee's account \\
      & payee's identifier & balance \\
      \hline
      Role of intermediary in  & update of asset ownership can only be  & account holders send signed \\
      transaction processing & initiated by the owner with his/her  & instructions to the intermediary;   \\
      & digital signature; intermediary only  & intermediary is responsible \\
      & verifies whether the asset has been & for updating account balances \\
      & previously consumed & \\
      \hline
      Required trust on intermediary  & Intermediary is only trusted to   & Intermediary is trusted to maintain  \\
      by asset issuers & maintain a unique, temporal order   & correct account balances of assets; \\
      & of transactions; intermediary is unable    & intermediary is able to issue new \\
      & to issue new units of assets as  & units of assets by modifying the \\
      & coinbase transactions for issuing new  & balance of an account holder\\
      & units require an issuer signature & \\
      \hline
      Double spending fraud  & Yes, an owner can spend the same & No, an account holder's balance is \\
      (by owner) & unit of asset multiple times to pay & debited multiple times accordingly \\
      & different  payees & if he sends in  multiple transactions \\
      & & paying different payees\\
      \hline
      Replay attack fraud  & No, no change in the final system  & Yes, the respective account balance  \\
      (by intermediary) & state if intermediary replays a & will be debited multiple times  \\
      & transaction multiple times & if intermediary replays a transaction \\
      & & multiple times\\
      \hline
      Transaction traceability & Yes & No\\
      \hline
    \end{tabular}
    }
    \vskip 0.5cm
    \caption{Differences between UTXO-based and account-based systems.}
    \label{table:utxo_versus_account}
\end{table}

\subsection{Differences between UTXO-based and token-based systems.}
Although both UTXO-based systems and pure token-based systems use the same type of global state representations, there are two subtle difference due to different designs of the state transaction operations, and due to the need of a record system (in physical existence) in the digital domain.

First, while the value of a payment object in a pure token-based system stays the same when transfer of ownership from one owner to another occurs, UTXO-based implementations usually allow splitting and merging to occur in transfer, which may results in a change of value of a resulting UTXO and a change in the total number of UTXOs in the global state of the system.  In formal definition, the processing of a splitting transaction and a merging transaction (for P2PKH transactions) is given as follows:
\[
\text{Splitting: } \qquad \qquad {\cal S}_{t+1}^{UTXO} = \{{\cal S}_t^{UTXO} \backslash \{(o_k, u_i)\}\} \bigcup \{ (o_x, u_j), (o_y, u_i)\}
\]
\[
\text{Merging: } \qquad \qquad {\cal S}_{t+1}^{UTXO} = \{{\cal S}_t^{UTXO} \backslash \{(o_k, u_i), (o_r, u_i)\}\} \bigcup \{ (o_x, u_j) \}
\]

In the splitting transaction, a user $u_i$ pays a portion of the value of his UTXO $o_k$ to another user $u_j$ resulting in a new UTXO $o_x$ under the ownership of $u_j$ and a change to himself resulting in another new UTXO $o_y$ under his ownership.  In the merging transaction, a user $u_i$ pays two UTXOs $o_k, o_r$ under his ownership to another user $u_j$, merging them into one new object $o_x$.

Second, due to the need of a record system, be it centralised or distributed in a blockchain, to record all UTXO transactions ever happened.  Anyone can easily chain up transactions together to form an inheritance chain of UTXOs since each transaction explicitly specify the end state of processing it.  As a consequence, the history of the ownerships of a given UTXO can be easily discovered, leading to transaction traceability in UTXO-based system.  This traceability is not easily done in account-based systems since each transaction is not self-contained with explicit end state information; one has to combine the system state with transaction to trace the history of funds which is time consuming.  This transaction traceability of Bitcoin or digital tokens is contrary to the conventional understanding that tokens are not traceable.  For physical token-based systems like cash and commodity money, such a record system does not physically exists; when a payment object changes hands, the old record would be erased and replaced with the new record.  Hence, it is not possible to chain up the history of ownership of a payment object.  This is the difference between physical tokens and UTXOs.

\subsection{A New Taxonomy}
\label{ssect:new_taxonomu}

As can be seen in Table~\ref{table:utxo_versus_account}, the difference in data structure used in the record systems leads to different capabilities and risks between a UTXO-based system and an account-based system.  It is natural to ponder if there is a need to classify the two types of arrangements in different categories.  Yet, UTXO-based systems are also considerably different from pure token-based systems.  A new taxonomy could be possible if the data structure used to represent the global state of the system is used as a defining feature to distinguish between token-based and account-based systems, as depicted in Table~\ref{table_taxonomy}.  That is, a token-based system is defined as one with an object-oriented global/system state representation, whereas, an account-based system is defined as one with an account-oriented global/system state representation.

\begin{table}[hbtp]
    \begin{center}
        \includegraphics[width=17.5cm]{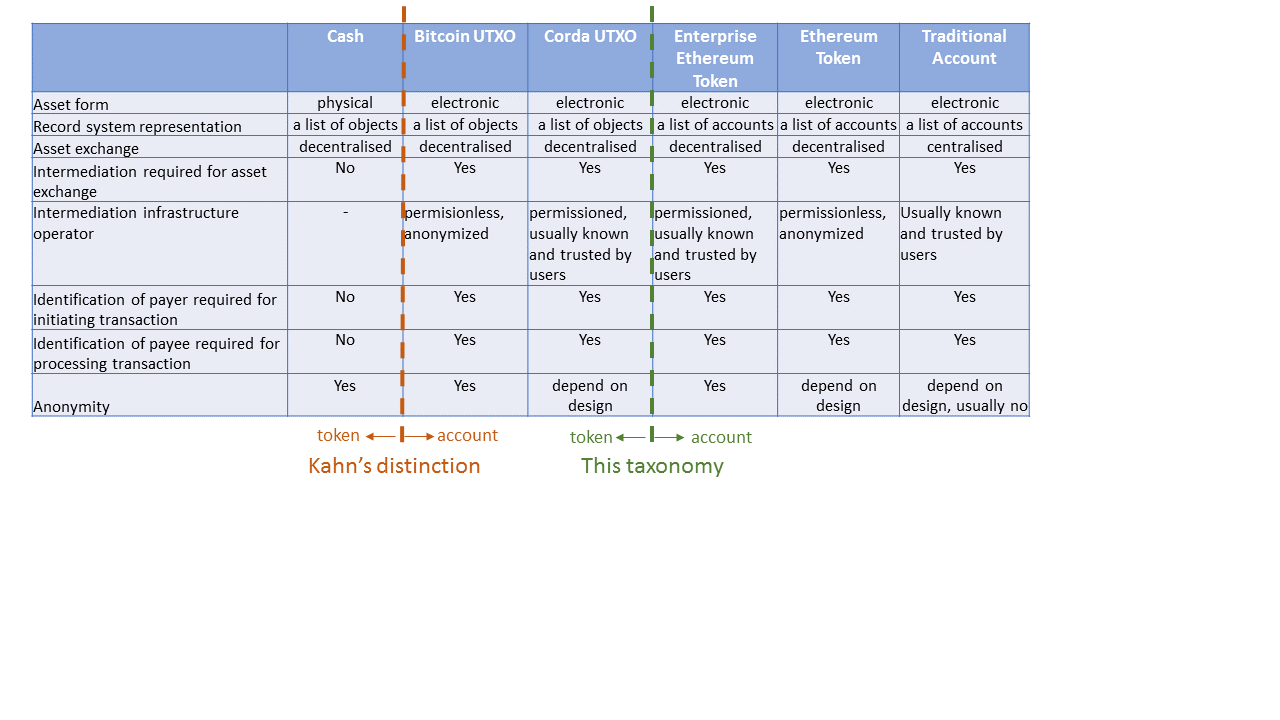}
    \end{center}
    \vskip -3.5cm
  \caption{A Taxonomy for digital tokens, compared with other instruments}
  \label{table_taxonomy}
\end{table}

\begin{figure}
  \centering
  \includegraphics[width=15cm]{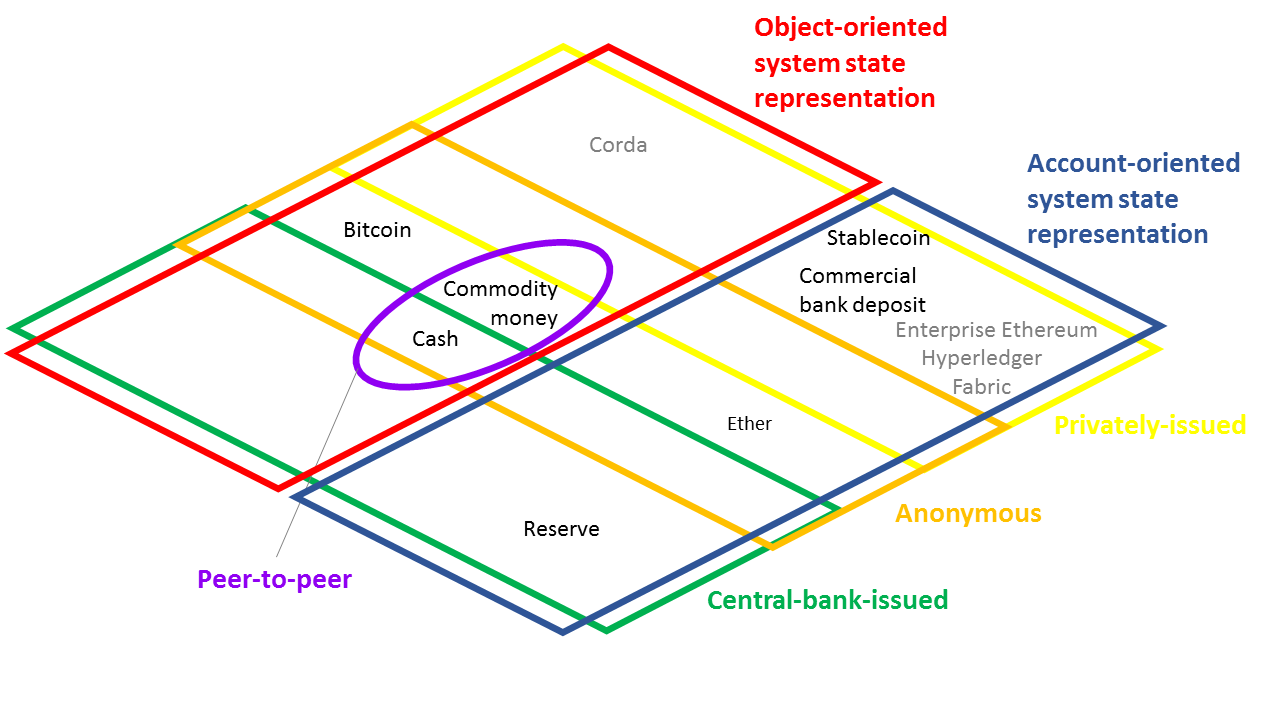}
  \vskip -1cm
  \caption{Money diamond: A new taxonomy for money}\label{figure_money_diamond}
\end{figure}

A Venn diagram, called a money diamond, is shown in Figure~\ref{figure_money_diamond} to present a taxonomy of money based on this distinction between token-based systems and account-based systems.  In this taxonomy, a token-based system records the state of the system as a list of individual assets or tokens or UTXOs, marked with the identifier of the corresponding owner who can spend the asset.  Note that the identifier could be anonymous.  Whereas, an account-based system records the state of the system as a list of accounts and their balances.  This definition extends the economic notion of token-based systems \cite{KahnR09} to include UTXO-based systems, such as Bitcoin.  Whether a record system or intermediation is required is not part of the defining features of this definition.  Token-based systems that do not require intermediation, that is, support peer-to-peer transfer of ownership, is subset, referring to physical token-based system such as cash and commodity money (e.g. gold, silver).

A payment instrument could be issued by a central bank, issued privately (by a commercial bank or payment service provider), or issued by nobody but the software platform (such as Bitcoin and Ether\footnote{Ether is the native cryptocurrency on Ethereum.}).  Note that common distributed ledger platforms (such as Corda, Hyperledger Fabric and Enterprise Ethereum) are included in the money diamond to illustrate the distinction between token-based systems and account-based systems in the digital domain.  But these platforms are not money; they could be used to issue money.  While stablecoins could be issued based on a token-based system or an account-based system, it seems that the underlying blockchain platforms of all of the existing stablecoins (such as Diem) are account-based.

The benefit of this classification or taxonomy is that physical tokens (such as cash and commodity money) fulfill this definition of tokens and are a subset of the token category while UTXO-based digital tokens can still be covered.  Besides, this taxonomy would not mistakenly classify conventional account-based systems (such as online bank accounts and RTGS systems) as token-based.  The flip side is that cryptocurrencies on Ethereum would be classified as account-based, which seems reasonable.  The feature of supporting anonymity should not be seen as a defining feature for tokens because supporting anonymity and token-based implementations are independent.  Supporting anonymity is a design choice of the underlying record system: whether it is permissionless and run by anonymous parties (and therefore cannot be held accountable for identify users), and whether user identification is part of the requirements for using services provided by the record system.

\section{Conclusion}
\label{sect::conclusion}
The term ``tokens'' is widely used in the discussion of digital currencies and crypto-assets but subject to different interpretations.  On the other hand, the distinction between token-based and account-based systems is well established in payment economics.  Through a detailed exposition of the design of UTXO-based systems such as Bitcoin, this article discusses why UTXO-based systems should be seen as account-based systems.  Understanding this reality would have practical implications on anonymity and system interoperability.  In addition, a comparison of UTXO-based systems and account-based systems is given, with a discussion on a new taxonomy which classifies UTXO-based systems as token-based systems.  The proposed definition of token-based systems based on their global/system state representation, regardless of whether a record system is required, is an extension of the classical economic notion of tokens which covers both physical and digital tokens while neatly distinguishing token-based and account-based systems.

\bibliographystyle{plain}
\bibliography{./token_account}

\end{document}